\documentclass[sigconf]{acmart}
\AtBeginDocument{%
  \providecommand\BibTeX{{%
    \normalfont B\kern-0.5em{\scshape i\kern-0.25em b}\kern-0.8em\TeX}}}

\setcopyright{acmcopyright}
\copyrightyear{2025}
\acmYear{2025}
\acmDOI{XXXXXXX.XXXXXXX}

\acmConference[Conference acronym 'XX]{Make sure to enter the correct
  conference title from your rights confirmation email}{June 03--05,
  2018}{Woodstock, NY}
\acmPrice{15.00}
\acmISBN{978-1-4503-XXXX-X/18/06}

\usepackage{hyperref}
\usepackage{cleveref}
\usepackage{subcaption}
\usepackage{enumitem}
\usepackage{algorithm}
\usepackage{algpseudocode}
\usepackage{amsmath}
\usepackage{bbm}
\begin{document}

\title{ReAlign: Reward-Guided Preference Alignment for Generative Re-Ranking}

\author{YaChen Yan}
\email{yachen_yan@intuit.com}
\orcid{0000-0002-1213-4343}
\affiliation{%
    \institution{Intuit Credit Karma}
    \streetaddress{760 Market Street}
    \city{Mountain View}
    \state{California}
    \country{USA}
    \postcode{94102}
}

\author{Liubo Li}
\email{liubo_li@intuit.com}
\orcid{1234-5678-9012}
\affiliation{%
    \institution{Intuit Credit Karma}
    \streetaddress{760 Market Street}
    \city{Mountain View}
    \state{California}
    \country{USA}
    \postcode{94102}
}

\renewcommand{\shortauthors}{YaChen and LiuBo}
\renewcommand{\subtitle}{ReAlign: Reward-Guided Preference Alignment for Generative Re-Ranking}


\begin{abstract}
Re-ranking refines the upstream ranked list to maximize listwise user utility. Existing two-stage generator-evaluator methods decouple list generation from list evaluation, leaving the generator without direct access to the listwise utility signal, while scoring-based one-stage methods score items under a fixed upstream permutation context without explicitly searching the permutation space. We propose \textbf{ReAlign}, a reward-guided preference alignment framework that couples a listwise reward model with a candidate-aware autoregressive generator. The reward model decomposes the listwise reward into position-wise contributions, modeling user exposure through a survival-based exposure module and item value through per-objective heads for heterogeneous engagement signals. The generator is trained in two stages: supervised pre-training on logged exposure lists to establish a relevance prior, followed by reward-guided post-training that aligns the generator with the reward model through complementary distribution and ranking alignment over within-pool standardized rewards. Extensive offline experiments show that ReAlign consistently outperforms existing state-of-the-art baselines.
\end{abstract}



\maketitle

\section{Introduction}

Re-ranking is the final stage of a modern recommender system, where an upstream ranking module produces a small candidate set and the re-ranker arranges these candidates into an ordered list shown to the user. Although the candidate set is short, the number of possible orderings grows factorially, and the user utility of a list depends on intra-list interactions such as competition, complementarity, and position-dependent exposure. Effectively searching this combinatorial space, while modeling how users actually consume an ordered list, is the central challenge of re-ranking.

Despite substantial progress, two gaps remain. First, listwise reward estimation is often coarse: existing evaluators commonly collapse user exposure and engagement into a single scalar, ignoring the cascading nature of user browsing and the heterogeneity of engagement objectives that arise in real systems. A re-ranker that relies on such a coarse signal cannot easily distinguish lists that differ only in deep-position behavior or in the trade-off across objectives. Second, even with a high-quality listwise signal, transferring that signal to a generator that explicitly searches the permutation space is non-trivial. Two-stage generator-evaluator pipelines decouple generation from evaluation and limit end-to-end optimization, while preference-based generator training is sensitive to the absolute scale of the reward and is known to be unstable when applied from random initialization in cold-start settings~\cite{lin2025gref}.

We propose \textbf{ReAlign}, a reward-guided preference alignment framework that addresses both gaps with a tightly coupled reward model and generator. The reward model decomposes the listwise reward into position-wise contributions, modeling user exposure as a discrete-time survival process and item value through multiple per-objective heads with uncertainty-based weighting. The generator is trained in two stages: a supervised pre-training stage that imitates logged exposure lists to establish a stable relevance prior, followed by a reward-guided post-training stage that aligns the generator with reward-induced listwise preferences over a diverse list pool. The post-training objective combines distribution alignment with a ranking alignment loss, both operating on within-pool standardized rewards to remain robust to the absolute scale and bounded biases of the reward.

Our contributions are summarized as follows.
\begin{itemize}
    \item We propose ReAlign, a reward-guided preference alignment framework for generative re-ranking that combines a listwise reward model with a candidate-aware autoregressive generator under a unified training pipeline.
    
    \item We design a decomposable listwise reward model that jointly models user exposure via a discrete-time survival process and item value via multi-objective heads with uncertainty-based weighting, yielding a calibrated and interpretable listwise reward.
    
    \item We develop a reward-guided post-training objective that combines distribution alignment and ranking alignment over a diverse list pool, with within-pool reward standardization that makes both terms robust to reward scale and bounded biases.

\end{itemize}

\section{Related Work}
\subsection{Re-ranking in Recommender Systems}
Re-ranking models aim to refine an initial ranked list by modeling item interactions and optimizing list-level user utility. Existing approaches can be broadly categorized into one-stage and two-stage methods.

One-stage methods directly model the target ranking in a single framework. Early works focus on listwise scoring with context modeling, where DLCM and PRM incorporate list context into deep models to refine item scores~\cite{ai2018learning, pei2019personalized}, and SetRank introduces permutation-invariant representations for set-level structure~\cite{pang2020setrank}. Some approaches directly model permutations as structured outputs emphasizing global optimization~\cite{zhuang2018globally, feng2021revisit}, while more recent work extends this to generative re-ranking in autoregressive or non-autoregressive manners~\cite{feng2021grn, ren2024non}.

Two-stage methods decompose re-ranking into a generator-evaluator framework. Seq2Slate generates slates via sequential decoding~\cite{bello2018seq2slate}, and PIER combines permutation-level generation with interest-aware evaluation~\cite{shi2023pier}. Recent works further improve candidate diversity and evaluation quality through multi-generator construction~\cite{yang2025comprehensive}, neighbor-aware generation~\cite{wang2025nlgr}, and unified generative frameworks~\cite{lin2025gref}. Newer studies also explore better optimization targets for the evaluator, including list-level value estimation and group-relative optimization~\cite{zhang2025generation, zhang2025goalrank}.
\subsection{Generative Recommendation and Preference Optimization}
Generative recommendation reframes item retrieval as sequence decoding. TIGER~\cite{rajput2023recommender} introduced semantic item IDs for autoregressive generative retrieval, and HSTU~\cite{zhai2024actions} scaled this paradigm to trillion parameters with LLM-like scaling behavior. Industrial deployments have followed~\cite{han2025mtgr, zhang2025gpr, ju2025generative, yang2025sparse}, and several works pursue end-to-end unification of retrieval and ranking: OneRec~\cite{deng2025onerec, zhou2025onerec} collapses the multi-stage pipeline into a single generative model; OneSearch~\cite{chen2025onesearch} applies this to e-commerce search; and GR4AD~\cite{xue2026generative} co-designs architecture, training, and serving for large-scale advertising.

Aligning generative recommenders with user preferences draws from RLHF and direct preference optimization. A related line of work extends preference contrast to listwise settings: PRO~\cite{song2024preference} decomposes a reward-ranked list into iterative one-vs-rest softmax classifications, and LiPO~\cite{liu2025lipo} connects preference optimization to learning-to-rank, improving alignment calibration over standard DPO-style objectives. In the recommendation context, OneRec~\cite{deng2025onerec, zhou2025onerec} applies iterative DPO alignment with online feedback; OneSearch~\cite{chen2025onesearch} employs a listwise DPO objective with adaptive reward-weighted ranking; and GR4AD~\cite{xue2026generative} proposes RSPO, an NDCG-weighted DPO variant that optimizes ranking quality beyond short-term click feedback. This line of work highlights that the choice of preference optimization objective plays a central role in enabling generative recommenders to align with real-world objectives.

\section{Methodology}
We describe ReAlign in detail, covering the framework overview, the shared user and item encoders, the listwise reward model, the candidate-aware generator with its two-stage training pipeline, and the overall training and inference procedure.

\subsection{Framework Overview}

Our framework, \textbf{ReAlign} (Re-ranking via reward-guided alignment), consists of two main components: a reward model and a generator. The reward model acts as an evaluator that estimates the expected utility of an ordered list by jointly modeling exposure and item value under user browsing behavior. The generator is a candidate-aware encoder--decoder model that autoregressively constructs ordered lists from the upstream candidate set. We train the framework in a staged manner. First, the reward model is learned from logged recommendation data with observed exposure and engagement signals. Then, the generator is trained in two stages: supervised pre-training on logged exposure lists, followed by reward-guided post-training through preference alignment. At serving time, the generator performs autoregressive decoding over the candidate set to produce the final re-ranked list.

\subsection{Problem Formulation}
We consider a multi-stage recommendation setting with a user space $\mathcal{U}$ and an item space $\mathcal{I}$.  
Let $u \in \mathcal{U}$ denote a user, and let $I \in \mathcal{I}$ denote an item.

Given a user $u$, an upstream ranking stage produces a set of candidate items $X = \{ I_1, I_2, \dots, I_m \}$, where $I_i \in \mathcal{I}$.

The re-ranking task aims to generate an ordered list of items $L = ( I_{l_1}, I_{l_2}, \dots, I_{l_n} )$, where $L \subseteq X$ and $n \ll m$.

We model re-ranking as a sequence generation problem. Specifically, we define a conditional distribution $P(L \mid u, X)$, where $L$ belongs to the space of ordered subsets of $X$. The ordered list is generated in an autoregressive manner:
\begin{equation}
P(L \mid u, X) = \prod_{t=1}^{n} P(I_{l_t} \mid u, X, I_{l_{<t}}).
\end{equation}

To evaluate and guide the generation of ordered lists, we define a reward function $r(u, L)$ over ordered lists conditioned on the user, which assigns a scalar value to each ordered list $L$ for user $u$.

The objective of the re-ranking problem is to learn a conditional distribution $P(L \mid u, X)$ that produces ordered lists $L$ maximizing the reward $r(u, L)$.

\subsection{User and Item Encoders}
A key component of our framework is a shared representation scheme that maps users and items into embedding tokens for downstream modeling. We introduce a user encoder and an item encoder to transform each user and each item into dense representations, which can be viewed as user and item tokens. These tokens serve as a common interface for subsequent modules. In the following, we first describe the user encoder and then present the item encoder.

\subsubsection{User Encoder}
\begin{figure}[htpb]
    \centering
    \includegraphics[width=0.50\textwidth, height=0.25\textwidth]{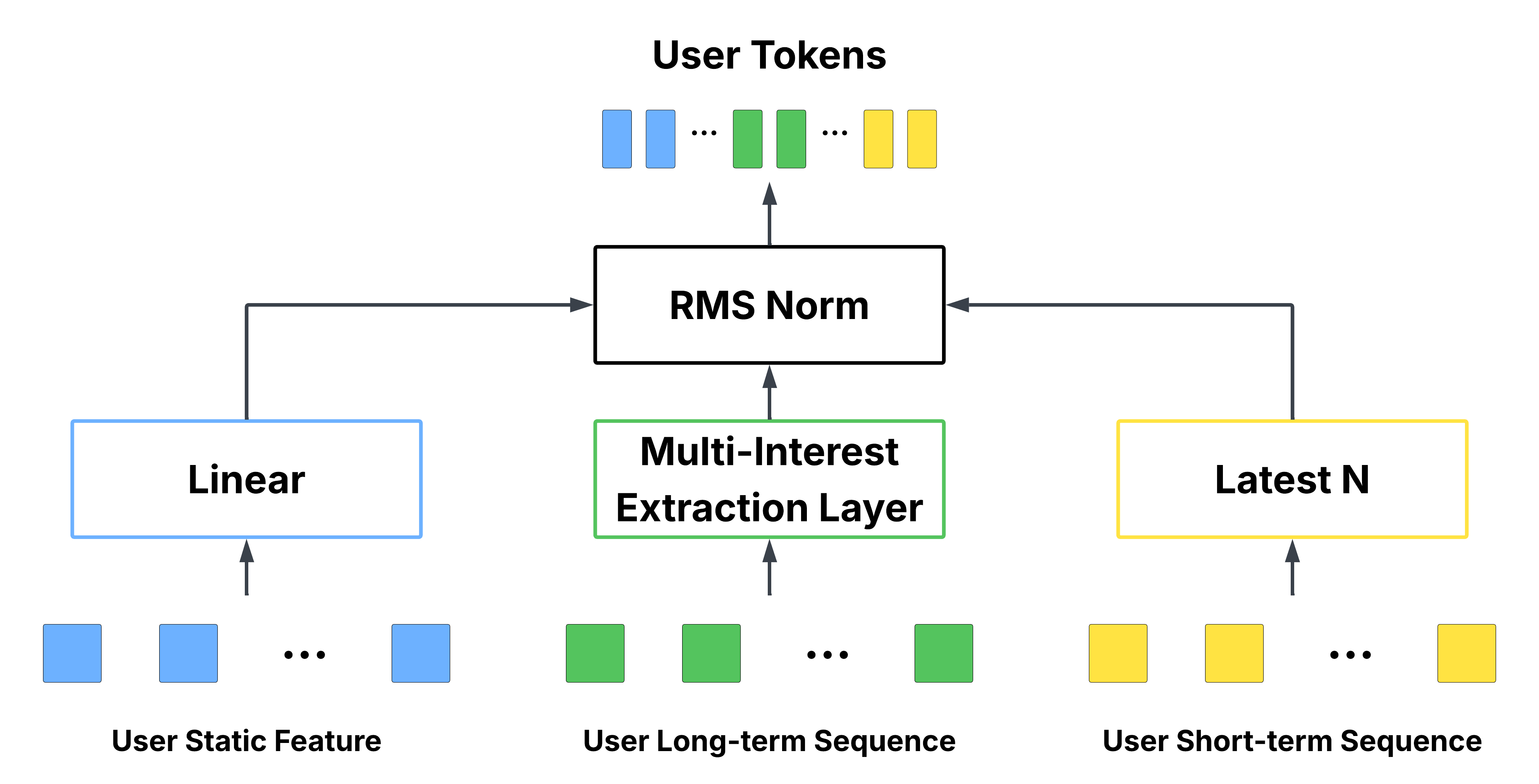}
    \caption{The user encoder maps heterogeneous user signals into a fixed set of user tokens, which are normalized to form a unified representation for downstream modules.}
    \label{fig_user_encoder}
    \medskip
    \small
\end{figure}

The user encoder transforms each user into a fixed set of user tokens that capture different aspects of user behavior. As illustrated in \autoref{fig_user_encoder}, it consists of three components that process user static features, long-term interaction sequences, and short-term interaction sequences, respectively.

For user static features, we apply a linear transformation to map the input features into a fixed-dimensional representation, which is then split into multiple tokens. For the long-term interaction sequence, we adopt a multi-interest extraction layer that extracts multiple interest vectors via a softmax-based attention mechanism~\cite{cen2020controllable}, each corresponding to a distinct latent interest. For the short-term interaction sequence, we use the most recent $N$ interactions as tokens to capture the user's immediate interests.
The tokens from these three components are concatenated to form the final set of user tokens, followed by RMS normalization to stabilize the representation.

\subsubsection{Item Encoder}
We adopt a lightweight design for the item encoder that combines item identity with side information. For each item, we learn a trainable ID embedding to capture collaborative signals, and encode its side information (e.g., category, price, and other metadata) into a dense representation via a linear projection. The two components are combined through element-wise addition, forming a unified item token that reflects both identity and feature semantics. Finally, we apply RMS normalization to stabilize the representation.

\subsection{Reward Model}
We introduce a reward model that plays the role of a listwise evaluator in the re-ranking stage, assigning a scalar reward to each candidate list for a given user. We use the term \emph{reward model} to emphasize its role in guiding downstream optimization, while it is conceptually equivalent to an evaluator in prior re-ranking frameworks.
 
Given a user $u$ and an ordered list of items $L = (I_{l_1}, \ldots, I_{l_n})$, the reward model is defined as $r_\phi(u, L)$, where $\phi$ denotes the model parameters.

\subsubsection{Listwise Reward Decomposition}
Rather than directly modeling the reward of the entire list, we decompose the listwise reward into position-wise contributions under the user browsing process:
\begin{equation}
\label{eq:reward_decomposition}
    r_\phi(u, L) = \sum_{t=1}^{n} \underbrace{p_\phi(t \mid u, L)}_{\text{reaching probability}} \cdot \underbrace{v_\phi(u, I_{l_t})}_{\text{item value}},
\end{equation}
where $p_\phi(t \mid u, L)$ denotes the probability that the user reaches and observes the item at position $t$, and $v_\phi(u, I_{l_t})$ denotes the expected value of the item conditional on being observed. Each item contributes to the listwise reward proportionally to the probability that the user actually sees it, yielding the expected utility under the browsing process. This decomposition naturally motivates modeling reaching probability and item value as two coupled components within a unified architecture.

\subsubsection{Reward Model Architecture}
\begin{figure}[htpb]
    \centering
    \includegraphics[width=0.50\textwidth, height=0.50\textwidth]{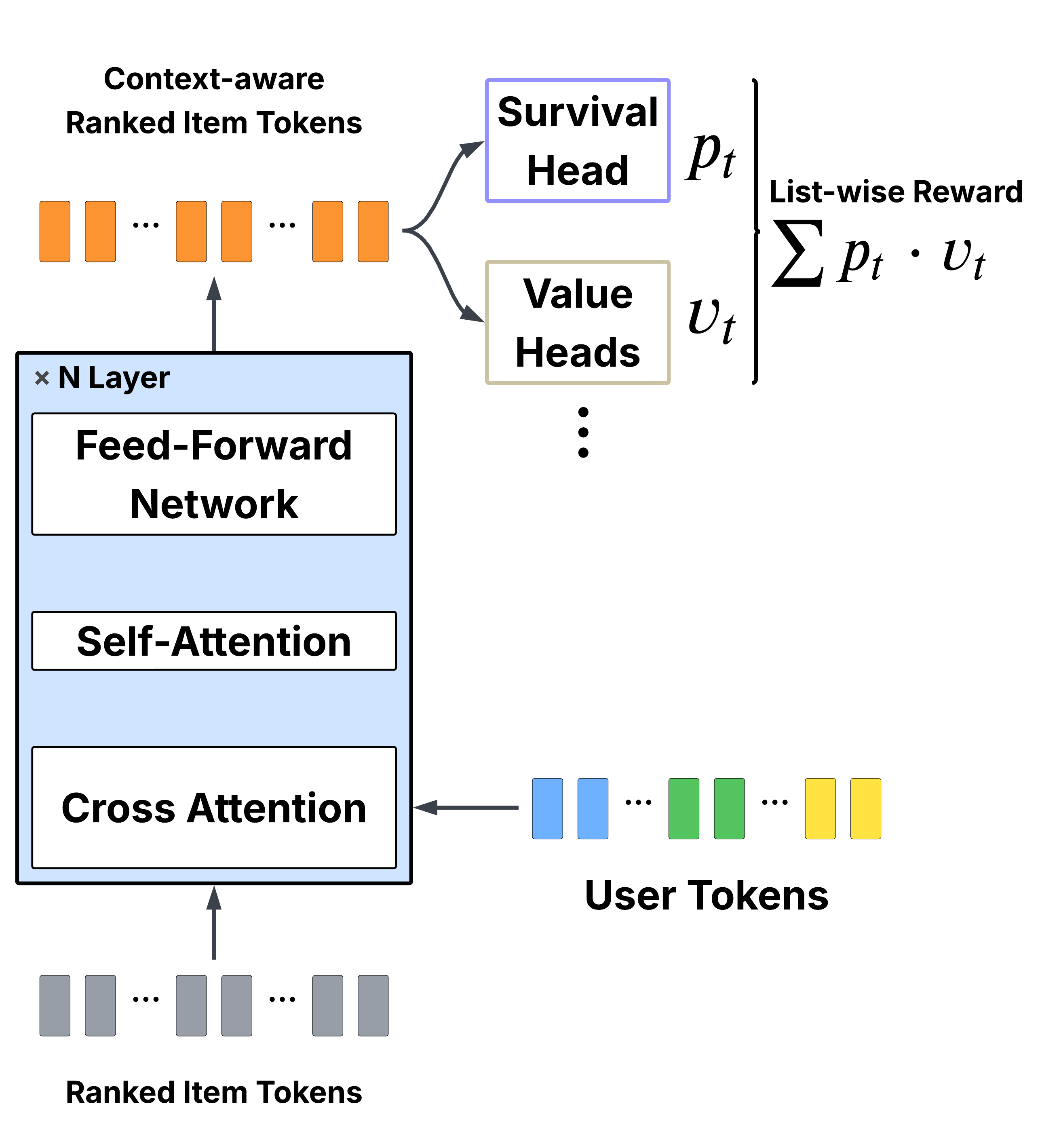}
    \caption{Architecture of the reward model. A transformer backbone encodes ranked item tokens via cross-attention to user tokens and self-attention over the ranked sequence, producing context-aware tokens that feed a survival head and $K$ value heads.}
    \label{fig_evaluator}
    \medskip
    \small
\end{figure}

To jointly model reaching probability and item value, we employ a unified transformer-based architecture over the ordered list, as illustrated in \autoref{fig_evaluator}. This architecture takes ranked item tokens and user tokens as input and produces context-aware ranked item tokens for all positions.

Specifically, before being processed by the transformer, each ranked item token is summed with a learned position embedding corresponding to its rank in the list. Position is a critical signal in re-ranking, as it directly governs user exposure and shapes how items are perceived in context. The position-augmented item tokens are then processed by a stack of transformer layers, where each layer consists of cross-attention to user tokens, self-attention over the ranked item sequence, and a feed-forward network, each sub-layer equipped with RMS normalization and residual connections. The cross-attention module injects user-specific preferences into each position, the self-attention module captures listwise dependencies among items (e.g., competition, redundancy, and complementarity), and the feed-forward network introduces position-wise nonlinearity to refine the contextualized representations.

The resulting context-aware ranked item tokens are then fed into two types of task-specific heads:
\begin{itemize}
    \item a \textbf{survival head} predicts the position-wise hazards $h_\phi(t \mid u, L)$, from which reaching probabilities can be derived,
    \item $K$ \textbf{value heads}, one per objective, each predicting a per-objective value estimate $\hat{v}_\phi^{(k)}(u, I_{l_t})$ that is later aggregated into the item value $v_\phi(u, I_{l_t})$.
\end{itemize}
This design disentangles reaching probability from item value while allowing joint optimization through shared representations.

\subsubsection{Exposure Modeling}
We model the user's exposure at each position via a \emph{reaching probability} $p_\phi(t \mid u, L)$, defined as the probability that the user reaches and observes the item at position $t$. Let $E_t \in \{0,1\}$ denote whether the item at position $t$ is observed. We assume a top-down sequential browsing process, where users scroll through the list and may exit at any position.

\vspace{0.5em}
\noindent\textbf{Sequential Browsing Intuition.}
In practice, users interact with a recommendation list sequentially: at each position, a user either exits or continues scrolling to the next. This means exposure is inherently cascading: a user cannot reach position $t$ without having chosen to continue past all preceding positions. Such behavior is naturally captured by a survival process, where each position presents a discrete risk of exit, and the probability of reaching deeper positions decreases monotonically as these risks accumulate.

\vspace{0.5em}
\noindent\textbf{Survival-Based Formulation.}
We formalize this intuition as a discrete-time survival process over a ranked list of $n$ positions. Let $M$ denote the position at which the user exits. The fundamental quantity is the \textbf{conditional hazard}:
\begin{equation}
    h_\phi(t \mid u, L) = P(M = t \mid M \geq t,\ u, L),
\end{equation}
the probability of exit at position $t$ given the user has reached it. At each position, the user exits with probability $h_\phi(t)$ or continues with probability $1 - h_\phi(t)$. The reaching probability is then obtained as the survival function of this process:
\begin{equation}
\label{eq:reaching_prob}
    p_\phi(t \mid u, L) = \prod_{j=1}^{t-1}\bigl(1 - h_\phi(j \mid u, L)\bigr),
\end{equation}
with $p_\phi(1 \mid u, L) = 1$ by convention, yielding a monotonically decreasing reaching probability with depth that recovers the cascading exposure structure.

\vspace{0.5em}
\noindent\textbf{Training Objective.}
Let $C = \min(M, n)$ denote the censoring position, and $\delta = \mathbf{1}[M \le n]$ the event indicator. The likelihood of an observed browsing trajectory under the discrete-time survival process is $P(C, \delta \mid u, L) = h_\phi(C \mid u, L)^{\delta} \prod_{t=1}^{C-\delta} \bigl(1 - h_\phi(t \mid u, L)\bigr)$, yielding the negative log-likelihood
\begin{equation}
\mathcal{L}_{\text{exp}} = -\,\delta \log h_\phi(C \mid u, L) \;-\; \sum_{t=1}^{C-\delta} \log\bigl(1 - h_\phi(t \mid u, L)\bigr).
\end{equation}
Equivalently, defining per-position exit indicators $y_t = \delta\,\mathbf{1}[t = C]$ over the reached positions $t \le C$, $\mathcal{L}_{\text{exp}}$ is the binary cross-entropy between $y_t$ and $h_\phi(t \mid u, L)$; right-censored trajectories ($\delta = 0$) contribute survival terms at every reached position and no event term.

\subsubsection{Value Modeling}
In practice, recommendation systems typically involve multiple engagement objectives, such as clicks, watch time, likes, and shares. Rather than collapsing these signals into a single scalar through a fixed weighted combination, we adopt a multi-objective framework that models each objective independently with a dedicated prediction head, improving both interpretability and optimization.

Formally, let $\mathcal{O} = \{1, \dots, K\}$ denote the set of $K$ objectives. For each objective $k \in \mathcal{O}$, let $Y_t^{(k)} \geq 0$ denote the observed utility for objective $k$ at position $t$, which is sparse and non-negative (e.g., zero for unengaged impressions and positive for engaged ones). The aggregated item value is defined as a weighted sum of per-objective value estimates:
\begin{equation}
\label{eq:item_value}
v_\phi(u, I_{l_t}) = \sum_{k=1}^{K} w_k \cdot \hat{v}_\phi^{(k)}(u, I_{l_t}),
\end{equation}
where $\hat{v}_\phi^{(k)}(u, I_{l_t})$ is the per-objective value estimate for objective $k$, and $w_k$ is the manually specified business importance weight that reflects the relative value of each objective at inference time. These weights encode product-level preferences and are distinct from the training-time loss weights introduced later for balancing optimization across heterogeneous objectives. 

\vspace{0.5em}
\noindent\textbf{Per-Objective Head and Weighted Logistic Regression.}
Each objective $k$ is modeled via a dedicated value head that produces a scalar logit $f_\phi^{(k)}(u, I_{l_t})$ from the shared context-aware representation. Following \citet{covington2016deep}, we adopt weighted logistic regression to handle sparse, non-negative utility labels. Every observed impression contributes a unit negative term, and an impression with $Y_t^{(k)} > 0$ additionally contributes a positive term weighted by $Y_t^{(k)}$, so exponentiating the logit recovers a calibrated estimate of the expected utility. At inference time, the per-objective value estimate is therefore obtained by exponentiating the logit:
\begin{equation}
\hat{v}_\phi^{(k)}(u, I_{l_t}) = \exp\!\left(f_\phi^{(k)}(u, I_{l_t})\right).
\end{equation}

\vspace{0.5em}
\noindent\textbf{Per-Objective Training Loss.}
The per-objective loss is a weighted cross-entropy over observed positions:
\begin{equation}
\mathcal{L}_{\text{value}}^{(k)} =
-\sum_{t:\, E_t = 1}
\left[
Y_t^{(k)} \cdot \log \sigma\!\left(f_\phi^{(k)}(u, I_{l_t})\right)
+ \log \left(1 - \sigma\!\left(f_\phi^{(k)}(u, I_{l_t})\right)\right)
\right],
\end{equation}
where $\sigma(\cdot)$ is the sigmoid function. Positive examples with larger $Y_t^{(k)}$ contribute proportionally more to the loss, guiding the model to accurately predict the magnitude of the utility rather than merely its occurrence.

\vspace{0.5em}
\noindent\textbf{Uncertainty-Based Multi-Task Weighting.}
A key challenge in multi-objective learning is balancing the training losses across objectives with different scales and noise levels, independently of the business importance weights used at inference time. Following \citet{kendall2018multi}, we adopt an uncertainty-based multi-task learning approach that automatically learns per-task loss weights by modeling the homoscedastic uncertainty of each objective. For each objective $k$, we introduce a learnable log-uncertainty parameter $s_k = \log \sigma_k^2$, where $\sigma_k > 0$ captures the observation noise of task $k$. The uncertainty-weighted loss for objective $k$ is: $\widetilde{\mathcal{L}}_{\text{value}}^{(k)} = e^{-s_k} \mathcal{L}_{\text{value}}^{(k)} + \frac{1}{2} s_k$.
The term $e^{-s_k}$ down-weights objectives with higher uncertainty, while the regularization term $\frac{1}{2} s_k$ prevents the model from trivially driving $\sigma_k \to \infty$. The parameters $\{s_k\}_{k=1}^{K}$ are optimized jointly with the model parameters $\phi$.

\vspace{0.5em}
\noindent\textbf{Training Objective.}
The overall value training objective is:
\begin{equation}
\mathcal{L}_{\text{value}} = \sum_{k=1}^{K} \widetilde{\mathcal{L}}_{\text{value}}^{(k)}
= \sum_{k=1}^{K} \left( e^{-s_k} \mathcal{L}_{\text{value}}^{(k)} + \frac{1}{2} s_k \right).
\end{equation}

\subsubsection{Joint Training Objective}
The reward model can be viewed as factorizing a joint distribution over exposure and engagement signals:
\begin{equation}
P(E_{1:n}, \mathbf{Y}_{1:n} \mid u, L) = P(E_{1:n} \mid u, L) \cdot P(\mathbf{Y}_{1:n} \mid E_{1:n}, u, L),
\end{equation}
where $\mathbf{Y}_{1:n} = \{Y_t^{(k)}\}_{t=1,\dots,n;\, k=1,\dots,K}$ denotes the collection of per-position multi-objective engagement signals, and the reward model is trained by minimizing:
\begin{equation}
\mathcal{L} = \mathcal{L}_{\text{exp}} + \mathcal{L}_{\text{value}}.
\end{equation}
At inference time, the listwise reward in Eq.~\eqref{eq:reward_decomposition} is computed by combining the reaching probability (Eq.~\eqref{eq:reaching_prob}) with the aggregated item value (Eq.~\eqref{eq:item_value}), yielding a list-level utility that jointly reflects exposure probability, item value and inter-item dependencies induced by the ordering.

\subsection{Generator}
The generator constructs ordered lists by autoregressively selecting items from the upstream candidate set. This candidate-aware design substantially reduces the search space while preserving sequential dependencies. The reward model provides training signals that guide the generator toward producing high-quality lists.

\subsubsection{Generator Model Architecture}
\begin{figure}[htpb]
    \centering
    \includegraphics[width=0.50\textwidth, height=0.45\textwidth]{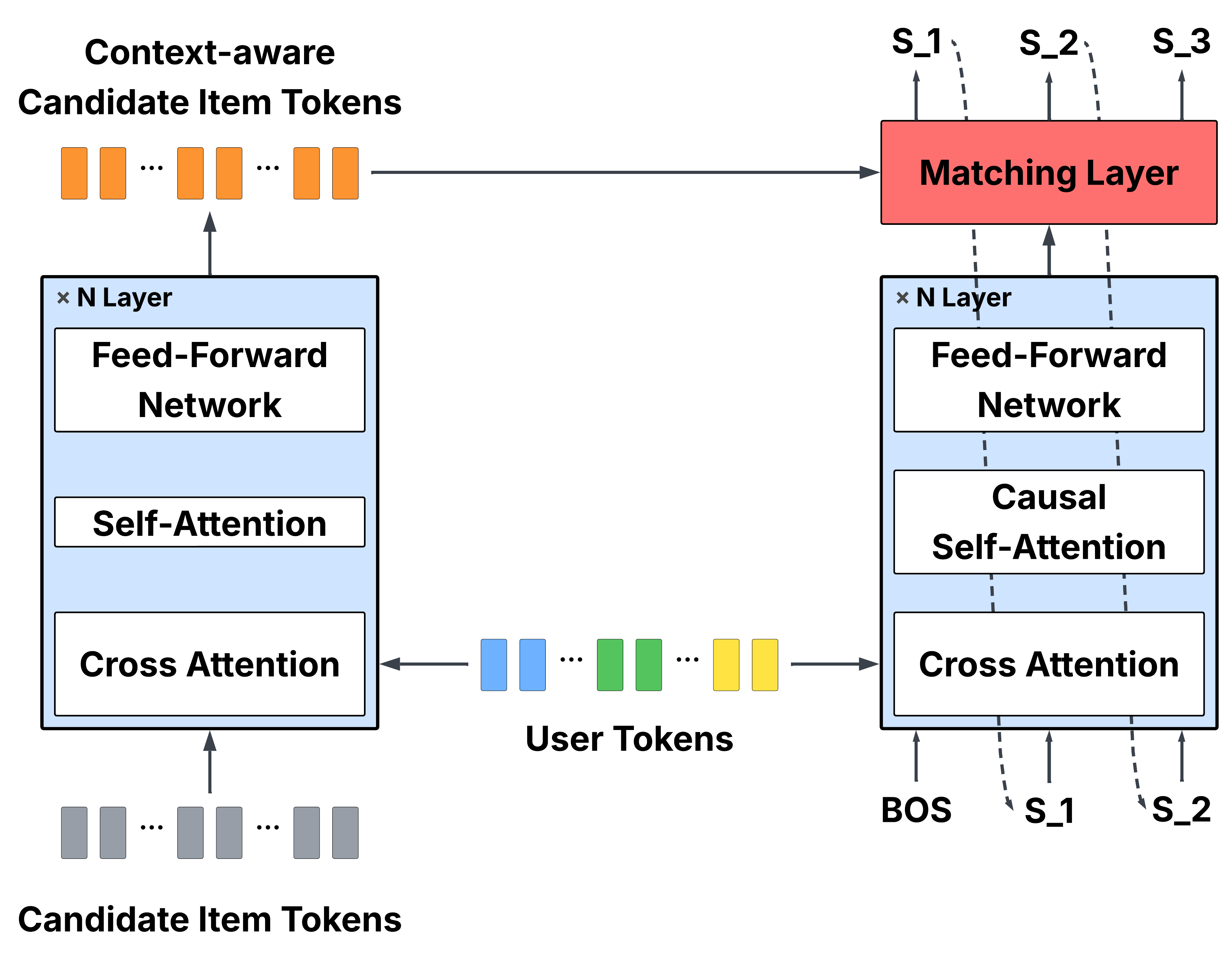}
    \caption{Architecture of the generator. Candidate items are encoded into context-aware item tokens, and a decoder autoregressively generates the re-ranked list. At each step, a matching layer selects the next item from the candidate set based on the decoder's query, with masking to avoid reselection.}
    \label{fig_generator}
    \medskip
    \small
\end{figure}

The generator follows an encoder--decoder architecture, as illustrated in \autoref{fig_generator}. Given a user $u$ and a candidate set $X=\{I_1,\dots,I_m\}$, the encoder first transforms candidate item tokens into contextualized candidate representations. This encoder is largely the same as that used in the reward model, except that here it operates on candidate item tokens rather than ranked item tokens. In this way, the encoder summarizes the candidate pool under the user condition and provides a dynamic, context-dependent item vocabulary for generation.

The decoder generates the ranked list autoregressively. At generation step $t$, it models the conditional distribution $P_\theta(I_{l_t}\mid u, X, I_{l_{<t}})$ over the candidate set. Each decoder layer consists of cross-attention to the encoded candidate representations and user tokens, causal self-attention over previously selected items, and a feed-forward network. Cross-attention injects user-specific and candidate-aware context into the decoding process, while causal self-attention captures dependencies among previously selected items and enforces sequential generation.

A key component of the generator is the matching-based selection mechanism. Instead of predicting over the full item vocabulary, the decoder produces a step-specific query representation $q_t$ at each position, which serves as a query for selecting the next item from the candidate set. This representation is then matched against the contextualized candidate item tokens $\{z_1,\dots,z_m\}$ generated by the encoder.

Specifically, for each candidate item $I_j$ that has not yet been selected, we compute a matching score:
\begin{equation}
s_t(j)=q_t^\top z_j, \quad I_j \in X_{\mathrm{rem}}^{(t)},
\end{equation}
where $X_{\mathrm{rem}}^{(t)}$ denotes the remaining candidate set at step $t$. The probability of selecting the next item is given by:
\begin{equation}
P_\theta(I_j\mid u,X,I_{l_{<t}})
=
\frac{\exp(s_t(j))}
{\sum_{I_{j'}\in X_{\mathrm{rem}}^{(t)}}\exp(s_t(j'))}.
\end{equation}
The next item is selected according to this distribution and removed from the candidate set via masking, ensuring that it will not be selected again. This process is repeated until the full ranked list is generated.

This matching-based decoding strategy enables efficient generation by restricting the action space to the candidate set, while still capturing sequential dependencies and listwise interactions through a dynamic, context-dependent item vocabulary induced by the encoded candidate representations.

\subsubsection{Pre-training on Logged Exposure Data}
We first pre-train the generator via supervised learning to provide a stable relevance prior before reward-guided post-training. Specifically, we use logged exposure data from the recommendation system, where each training instance consists of a user $u$, a candidate set $X$, and a historical exposed ordered list
\[
L = (I_{l_1}, I_{l_2}, \dots, I_{l_n}).
\]
The pre-training objective is to maximize the likelihood of the observed list under the autoregressive generator:
\begin{equation}
\mathcal{L}_{\mathrm{pre}}
=
- \log P_\theta(L \mid u, X)
=
- \sum_{t=1}^{n}
\log P_\theta(I_{l_t} \mid u, X, I_{l_{<t}}),
\end{equation}
where $\theta$ denotes the generator parameters.

This stage trains the generator to imitate historically exposed ordered lists under the candidate-aware generation setting, enabling it to learn a strong relevance prior and the sequential structure of recommendation lists from logged data. The resulting pre-trained generator provides a strong initialization for the subsequent reward-guided post-training stage.

\subsubsection{Post-training via Reward-Guided Preference Alignment}
After pre-training, we further refine the generator through reward-guided preference alignment. The key idea is to construct a diverse pool of candidate lists for each training instance, score them with the pretrained reward model, and then optimize the generator to align with the reward-induced listwise supervision.

\vspace{0.5em}
\noindent\textbf{List Pool Construction.}
For each training instance with user $u$ and candidate set $X$, we construct a list pool
\[
\mathcal{P}(u, X) = \{L^{(1)}, L^{(2)}, \dots, L^{(B)}\},
\]
where each $L^{(b)}$ is an ordered list over $X$ and $B = |\mathcal{P}(u, X)|$
denotes the pool size. To ensure sufficient diversity, the pool is constructed
from multiple sources, including historical exposed ordered lists, sampled
lists from the current generator under different decoding strategies, and
lists produced by alternative re-ranking strategies. This design mitigates
self-confirmation bias and exposes the generator to a broader set of plausible
ranking behaviors.

\vspace{0.5em}
\noindent\textbf{Reward-Based Supervision.}
Each list $L^{(b)} \in \mathcal{P}(u, X)$ in the pool is evaluated by the
pretrained reward model to obtain a score $r_\phi(u, L^{(b)})$. From these scores, we derive two complementary supervision signals: a reward-induced distribution over the pool and a reward-induced ordering over the pool.

\vspace{0.5em}
\noindent\textbf{Distribution Alignment.}
We first convert reward scores into a soft target distribution by 
standardizing them within the pool. Let
\begin{equation}
\label{eq:standardized_reward}
\tilde{r}_\phi(u, L^{(b)})
= \frac{r_\phi(u, L^{(b)}) - \bar{r}_\mathcal{P}}{\sigma_\mathcal{P} + \epsilon},
\end{equation}
where $\bar{r}_\mathcal{P} = \frac{1}{|\mathcal{P}(u, X)|} \sum_{L' \in \mathcal{P}(u, X)} r_\phi(u, L')$ and $\sigma_\mathcal{P}$ are the mean and standard deviation of $r_\phi(u, \cdot)$ over $\mathcal{P}(u, X)$, and $\epsilon > 0$ is a small constant for numerical stability. The reward-induced reference distribution is then the softmax of these standardized rewards:
\begin{equation}
\label{eq:pi_phi}\pi_{\phi}(L^{(b)} \mid u, X)
=
\frac{\exp\!\bigl( \tilde{r}_\phi(u, L^{(b)}) \bigr)}
{\sum_{L' \in \mathcal{P}(u, X)} \exp\!\bigl( \tilde{r}_\phi(u, L') \bigr)}.
\end{equation}
This construction follows the group-relative reference policy of GoalRank~\cite{zhang2025goalrank}: within-pool standardization makes $\pi_\phi$ robust to bounded biases in the reward model, since the induced ordering is preserved whenever reward gaps within $\mathcal{P}(u, X)$ are large relative to the bias. We adopt this principle as one of two complementary supervision signals, and further reuse the standardized rewards to modulate the ranking-alignment temperatures below.

To align the generator with this target, we define the generator-induced distribution over the same list pool:
\begin{equation}
\pi_{\theta}(L^{(b)} \mid u, X)
=
\frac{\exp\!\left(\log P_\theta(L^{(b)} \mid u, X)\right)}
{\sum_{L' \in \mathcal{P}(u, X)} \exp\!\left(\log P_\theta(L' \mid u, X)\right)}.
\end{equation}
We then optimize a cross-entropy loss:
\begin{equation}
\mathcal{L}_{\mathrm{dist}}
=
-\sum_{L \in \mathcal{P}(u, X)}
\pi_{\phi}(L \mid u, X)\log \pi_{\theta}(L \mid u, X).
\end{equation}

\vspace{0.5em}
\noindent\textbf{Ranking Alignment.}
Besides the distribution signal, we also enforce consistency between the ordering induced by the reward model and that induced by the generator over the same list pool. Following the Preference Ranking Optimization (PRO) framework~\cite{song2024preference}, we extend pair-wise contrast to an iterative one-vs-rest contrast over the entire ranking, with contrast temperatures modulated by the standardized reward gaps.

To keep the contrast strength comparable to $\mathcal{L}_{\mathrm{dist}}$ and independent of the absolute scale of $r_\phi$, we measure reward gaps in the same standardized space defined in Eq.~\eqref{eq:standardized_reward}. Without loss of generality, assume the lists in the pool are sorted in descending order of reward, so that $\tilde{r}_\phi(u, L^{(1)}) \geq \tilde{r}_\phi(u, L^{(2)}) \geq \cdots \geq
\tilde{r}_\phi(u, L^{(B)})$. At step $k$, we treat $L^{(k)}$ as the positive sample and $\{L^{(k')}\}_{k' > k}$ as negatives, contrasting them through a softmax over the generator log-likelihoods scaled by the corresponding reward gaps:
\begin{equation}
\label{eq:pro_step}
\mathcal{L}_{\mathrm{rank}}^{(k)} = -\log \frac{\exp\!\bigl( \log P_\theta(L^{(k)} \mid u, X) \big/ \mathcal{T}_{k,k} \bigr)
}{\sum_{k' = k}^{B} \exp\!\bigl( \log P_\theta(L^{(k')} \mid u, X) \big/ \mathcal{T}_{k,k'} \bigr)},
\end{equation}
where the per-pair temperature $\mathcal{T}_{k,k'}$ for $k' > k$ is determined by the standardized reward gap between $L^{(k)}$ and $L^{(k')}$:
\begin{equation}
\mathcal{T}_{k,k'} = 
\frac{1}{\max\bigl(\tilde{r}_\phi(u, L^{(k)}) - \tilde{r}_\phi(u, L^{(k')}),\ \epsilon\bigr)}, 
\qquad k' > k,
\end{equation}
with $\epsilon > 0$ a small constant that prevents division by zero when two lists have numerically indistinguishable standardized rewards. The positive sample's own temperature is set to the smallest among its paired temperatures, $\mathcal{T}_{k,k} = \min_{k' > k} \mathcal{T}_{k,k'}$, so that its log-likelihood is scaled with the sharpest factor within the step $k$. Larger standardized reward gaps therefore yield smaller temperatures and sharper contrasts, allocating more supervision capacity to pairs the reward model can confidently distinguish; measuring gaps in the standardized space $\tilde{r}_\phi$ additionally makes the temperature scale-free, adapting to within-pool reward dispersion rather than the absolute magnitude of $r_\phi$.

The overall ranking alignment loss aggregates contrasts across all positions in the ranking:
\begin{equation}
\mathcal{L}_{\mathrm{rank}} = \sum_{k=1}^{B-1} \mathcal{L}_{\mathrm{rank}}^{(k)}.
\end{equation}
By iteratively promoting the current top candidate against all lower-ranked ones, this objective progressively aligns the generator's likelihood ranking over $\mathcal{P}(u, X)$ with the ranking induced by the reward model.

\vspace{0.5em}
\noindent\textbf{Overall Objective.}
The final post-training objective is:
\begin{equation}
\mathcal{L}_{\mathrm{post}}
=
\mathcal{L}_{\mathrm{dist}}
+
\lambda_{\mathrm{rank}} \mathcal{L}_{\mathrm{rank}},
\end{equation}
where $\lambda_{\mathrm{rank}}$ controls the trade-off between distribution alignment and ranking consistency. This post-training stage refines the generator to better align with the reward-induced listwise preferences over the candidate pool.

\subsection{Training and Inference}

\subsubsection{Reward Model Training}

We train the reward model on logged recommendation data with observed exposure and engagement signals. The exposure component is learned from user browsing trajectories, while the value component is learned from item-level engagement signals on exposed positions. These two components are jointly optimized under the unified formulation to obtain a listwise reward model $r_\phi(u, L)$.

\subsubsection{Generator Training}

The generator is trained in two stages. In pre-training, we apply supervised learning on logged exposure lists, where the generator learns to model the conditional distribution $P_\theta(L \mid u, X)$ via autoregressive next-item prediction.

In post-training, we refine the generator through reward-guided preference alignment. For each $(u, X)$, we construct a list pool, score lists using the reward model, and optimize the generator to align with reward-induced preferences derived from the reward model.

\subsubsection{Generator Inference}

At serving time, only the generator is used; the reward model is needed only during training. Given user and candidate representations, the generator produces an ordered list autoregressively over the candidate set. At each step, the decoder's query representation $q_t$ is matched against the remaining candidate item tokens to compute the selection distribution, and the next item is selected greedily by taking the argmax, followed by masking to prevent reselection. We adopt greedy decoding for low-latency serving; beam search or stochastic sampling can be used when richer exploration is desired. This candidate-aware decoding enables efficient re-ranking without operating over the full item space.

\section{Experiments}
We evaluate ReAlign through two complementary studies: an offline comparison against strong baselines on two public benchmarks, and ablation studies validating each design choice. 

\subsection{Experimental Settings}

\subsubsection{Datasets}
We evaluate on two public datasets with complementary characteristics.
\begin{itemize}
    \item \textbf{KuaiRand}\footnote{\url{https://kuairand.com/}} is a large-scale randomly logged dataset from Kuaishou with multi-objective engagement signals (watch time, likes, follows, comments). Random item insertion yields unbiased exposure signals. The exit position is inferred as the first position after which all subsequent play times fall below a minimum threshold.
    \item \textbf{Avito}\footnote{\url{https://www.kaggle.com/c/avito-context-ad-clicks}} contains user search page logs with binary ad click signals. The exit position is inferred at the last clicked ad, with subsequent ads treated as unobserved.
\end{itemize}

\subsubsection{Evaluation Metrics}
We adopt two ranking metrics:
\begin{itemize}
    \item \textbf{NDCG} (Normalized Discounted Cumulative Gain) measures ranked list quality with a logarithmic position discount that rewards relevant items at higher positions.
    \item \textbf{uAUC} (User-level Generalized AUC) is the generalized AUC computed per user and averaged across users. Its generalized form supports non-binary engagement labels (e.g., watch time), and the per-user aggregation is more robust to user activity imbalance than global AUC.
\end{itemize}

\subsubsection{Baselines}
We compare against three categories: \textbf{pointwise models} that score items independently (DNN~\cite{covington2016deep}); \textbf{listwise contextual models} that capture mutual influence within a list (PRM~\cite{pei2019personalized}); and \textbf{permutation-level models} that evaluate ordered lists as a whole (PIER~\cite{shi2023pier}, CAVE~\cite{zhang2025generation}).

\subsubsection{Implementation Details}
All models are implemented in PyTorch. The user encoder extracts $8$ interest tokens from the long-term sequence and uses the most recent $N=20$ interactions as short-term tokens. The item encoder uses a $64$-dim ID embedding combined with side information projected to the same dimension via element-wise addition. The transformer backbone of both the reward model and generator has $3$ layers, $4$ attention heads, and hidden dimension $128$, with learned position embeddings matching the item token dimension. We set $K=1$ on Avito (click) and $K=4$ on KuaiRand (watch time, like, follow, comment), with uniform weights $\{w_k\}$ and log-uncertainty parameters $\{s_k\}$ initialized to $0$. Models are optimized with AdamW (learning rate $1\times 10^{-4}$, batch size $256$), with early stopping on validation NDCG (patience $3$).

For all evaluator-based approaches, we follow the MG-E protocol~\cite{yang2025comprehensive}: a fixed set of candidate permutations is scored by the reward model, and the highest-scoring permutation is selected.

\subsection{Overall Performance}
\autoref{tab_main} summarizes the offline performance on KuaiRand and Avito. For fair comparison with scoring-based re-rankers, we report two variants: \textbf{Evaluator Only} uses our reward model as a list scorer over candidate permutations, matching the inference protocol of the baselines; \textbf{ReAlign} is the full framework where the generator produces the final list after reward-guided preference alignment.

Two observations emerge. First, Evaluator Only already surpasses all baselines on both datasets and metrics, which we attribute to the survival-based exposure modeling and the disentangled multi-objective value estimation. Second, ReAlign further improves over Evaluator Only, showing that reward-guided preference alignment effectively transfers the reward model's signal to the generator.

\begin{table}[htpb]
\caption{Overall performance comparison on KuaiRand and Avito datasets.}
\label{tab_main}
\begin{tabular}{lcccc}
\toprule
  & \multicolumn{2}{c}{KuaiRand} 
  & \multicolumn{2}{c}{Avito} \\
\cmidrule(lr){2-3} \cmidrule(lr){4-5}
                 & NDCG            & uAUC            & NDCG            & uAUC            \\
\midrule
DNN              & 0.7965          & 0.6781          & 0.7022          & 0.6956          \\
PRM              & 0.8180          & 0.6973          & 0.7182          & 0.7019          \\
PIER             & 0.8312          & 0.7091          & 0.7281          & 0.7183          \\
CAVE             & 0.8450          & 0.7196          & 0.7372          & 0.7258          \\
\midrule
Evaluator Only   & 0.8523          & 0.7236          & 0.7450          & 0.7326          \\
ReAlign          & \textbf{0.8551} & \textbf{0.7258} & \textbf{0.7476} & \textbf{0.7342} \\
\bottomrule
\end{tabular}
\end{table}

\subsection{Ablation Study}
We conduct three sets of ablations to validate key design choices: (i) the reward model architecture, (ii) the post-training objective components, and (iii) the two-stage generator training pipeline. All experiments are conducted on KuaiRand, whose richer multi-objective signals make it more representative for examining fine-grained design choices.

\subsubsection{Reward Model Ablation}
We construct two variants targeting the core modeling components:
\begin{itemize}
    \item \textbf{w/o Survival} replaces the survival-based exposure module with a uniform reaching probability, removing position-dependent exposure discounting.
    \item \textbf{w/o Multi-Obj} collapses all engagement objectives into a single prediction head, removing both the per-objective heads and the uncertainty-based weighting.
\end{itemize}
\autoref{tab_ablation} shows that both variants degrade performance. Ablating the survival module yields the larger drop, confirming that position-dependent exposure discounting captures genuine user browsing behavior and is critical for accurate listwise reward estimation. Ablating the multi-objective design causes a smaller but consistent drop, showing that disentangling heterogeneous engagement signals further improves estimation beyond a single-objective formulation.

\begin{table}[htpb]
\caption{Ablation study of the reward model components.}
\label{tab_ablation}
\begin{tabular}{lcc}
\toprule
                  & NDCG            & uAUC            \\
\midrule
Full Reward Model & \textbf{0.8551} & \textbf{0.7258} \\
\midrule
w/o Survival      & 0.8341          & 0.7081          \\
w/o Multi-Obj     & 0.8487          & 0.7197          \\
\bottomrule
\end{tabular}
\end{table}

\subsubsection{Post-training Loss Ablation}
We compare the full loss against three variants:
\begin{itemize}
    \item \textbf{Only $\mathcal{L}_{\mathrm{dist}}$} drops the ranking alignment term, supervising the generator only through the reward-induced distribution.
    \item \textbf{Only $\mathcal{L}_{\mathrm{rank}}$} drops the distribution alignment term, supervising the generator only through the PRO-style one-to-$N$ contrast.
    \item \textbf{w/o Standardization} replaces the pool-standardized reward $\tilde{r}_\phi$ in Eq.~\eqref{eq:standardized_reward} with the raw reward $r_\phi$ in both $\mathcal{L}_{\mathrm{dist}}$ and the temperature of $\mathcal{L}_{\mathrm{rank}}$.
\end{itemize}
\autoref{tab_post_loss_ablation} shows that all three variants degrade performance. Removing within-pool standardization yields the largest drop, confirming that a pool-adaptive scale is critical: standardization keeps contrast strength comparable across the two terms and prevents either from being dominated by the absolute scale of $r_\phi$. Between the two loss components, dropping $\mathcal{L}_{\mathrm{dist}}$ degrades more than dropping $\mathcal{L}_{\mathrm{rank}}$, indicating that the reward-induced distribution carries the primary alignment signal, while the ranking term provides complementary sharpening of relative order. The two terms are nonetheless complementary, as the full objective outperforms both single-loss variants.

\begin{table}[htpb]
\caption{Ablation study of the post-training loss components.}
\label{tab_post_loss_ablation}
\begin{tabular}{lcc}
\toprule
                                          & NDCG            & uAUC            \\
\midrule
Full Post-training                        & \textbf{0.8551} & \textbf{0.7258} \\
\midrule
Only $\mathcal{L}_{\mathrm{dist}}$        & 0.8544          & 0.7247          \\
Only $\mathcal{L}_{\mathrm{rank}}$        & 0.8530          & 0.7242          \\
w/o Standardization                       & 0.8527          & 0.7233          \\
\bottomrule
\end{tabular}
\end{table}

\subsubsection{Training Stage Ablation}
We compare the full two-stage pipeline against two single-stage variants:
\begin{itemize}
    \item \textbf{Pre-training only} skips post-training and uses the pre-trained generator as the final model.
    \item \textbf{Post-training only} skips pre-training and applies reward-guided preference alignment from random initialization, with no relevance prior in place.
\end{itemize}
\autoref{tab_pipeline_ablation} shows that both single-stage variants underperform the full pipeline. Post-training alone reaches competitive performance and slightly outperforms pre-training only, but the full pipeline outperforms both, indicating that pre-training establishes a stable starting policy and post-training refines it to align with the reward-induced listwise preferences.

\begin{table}[htpb]
\caption{Ablation study of the training stage.}
\label{tab_pipeline_ablation}
\begin{tabular}{lcc}
\toprule
                                   & NDCG            & uAUC            \\
\midrule
Pre-training + Post-training       & \textbf{0.8551} & \textbf{0.7258} \\
\midrule
Pre-training only                  & 0.8305          & 0.7078          \\
Post-training only                 & 0.8369          & 0.7131          \\
\bottomrule
\end{tabular}
\end{table}




\section{Conclusion}

We proposed \textbf{ReAlign}, a reward-guided preference alignment framework for generative re-ranking. ReAlign couples a decomposable listwise reward model, based on survival exposure and multi-objective value estimation, with a candidate-aware autoregressive generator trained through supervised pre-training and reward-guided post-training. Extensive offline experiments on KuaiRand and Avito show that ReAlign consistently outperforms strong baselines, and ablation studies confirm the contribution of each design choice. Future work includes extending the reward model to delayed, long-horizon objectives and applying online policy distillation with token-level supervision to compress the generator into a lightweight student model for low-latency serving.


\newpage
\bibliographystyle{ACM-Reference-Format}
\bibliography{realign.bib}

\end{document}